\documentclass[aps,twocolumn,groupedaddress,floatfix]{revtex4-1}
\usepackage{amssymb,amsmath}
\usepackage{graphicx}
\usepackage{subfigure}
\usepackage[english]{babel} 
\usepackage{float}
\usepackage{color}

\usepackage{lipsum}
\begin{document}
%\DeclareRobustCommand{\baselinestretch{2}}
\newcommand{\be}{\begin{equation}}
\newcommand{\ee}{\end{equation}}
\newcommand{\rojo}[1]{\textcolor{red}{#1}}

\title{Fractional Bloch oscillations}

\author{Mario I. Molina}
\affiliation{Departamento de F\'{\i}sica, Facultad de Ciencias, Universidad de Chile, Casilla 653, Santiago, Chile}

\date{\today }

\begin{abstract}
We examine the effect of fractionality on the bloch oscillations (BO) of a 1D tight-binding lattice when the discrete Laplacian is replaced by its fractional form. We obtain the eigenmodes and the dynamic propagation of an initially localized excitation in closed form as a function of the fractional exponent and the strength of the external potential. We find an oscillation period  equal to that of the non-fractional case. The participation ratio is computed in closed form and it reveals that localization of the modes increases with a deviation from the standard case, and with an increase of the external constant field. When nonlinear effects are included, a competition between the tendency to Bloch oscillate, and the trapping tendency typical of the Kerr effect is observed, which ultimately obliterates the BO in the limit of large nonlinearity.

\end{abstract}

\maketitle

{\em Introduction}. Soon after the invention of calculus, some mathematicians wondered about a possible extension of Newton's calculus to non-integer derivatives. After some simple cases that made use of the Gamma function and the Fourier transform, it was shown that it 
could be done in principle for some simple cases. However, the idea of fractional calculus remained a mathematical curiosity until more recent times, when it was taken by mathematicians like Laplace, Euler, Riemann, and Caputo, to name a few, which promoted fractional calculus from a mathematical curiosity to a whole research field\cite{historical1,historical2,historical3,historical4,malomed}. Although classical physical systems are usually described by differential equations of integer order, the relevance of fractional order is found in its capacity to address complex systems with non-local or memory effects that are hard to treat with conventional methods. So far,  applications have been found in fluids mechanics\cite{fluids}, fractional kinetics and anomalous diffusion\cite{random walk, fractional kinetics, anomalous transport}, strange kinetics\cite{strange kinetics}, fractional quantum mechanics\cite{quantum1,quantum2}, Levy processes in quantum mechanics\cite{levy}, plasmas\cite{plasmas}, electrical conduction in cardiac tissue\cite{cardiac}, and biological invasions\cite{invasions}, among others.

Bloch oscillations (BO) are a quantum mechanical phenomenon where particles, such as electrons in a crystal lattice, undergo periodic motion when subjected to a constant external force, like an electric field. Predicted by Felix Bloch in 1929, these oscillations arise due to the wave-like nature of particles and the periodic potential of the lattice\cite{bloch}. Instead of accelerating indefinitely, as classical physics would suggest, the particle's motion becomes cyclic due to Bragg reflection at the edges of the Brillouin zone. While challenging to observe in natural solids due to scattering effects, Bloch oscillations have been experimentally confirmed in semiconductor superlattices\cite{wannier, superlattices1,superlattices2}, ultracold atoms\cite{ultracold1, ultracold2}, and photonic systems\cite{photonic systems 1, photonic systems 2}.

{\em The model}.\  Let us briefly review the standard Bloch oscillation phenomenon for completeness. Consider an electron or optical excitation propagating along a 1D tight-binding lattice in the presence of a constant external field. The dimensionless equations are
\be
i {d\over{d z}} a_{n}(z)+ n F a_{n}(z) + a_{n+1}(z)+a_{n-1}(z)=0\label{eq1}
\ee
where $a_{n}(z)$ is the electronic amplitude or the electric field amplitude in the optical case, $z$ is the propagation coordinate, and $F$ is a parameter proportional to the external constant field.
In the Fourier domain Eq.(\ref{eq1}) reads
\be
i {d\over{dz}} \tilde{a}(k) + i F {d\over{d k}}\tilde{a}(k) +  2\cos(k) 
\tilde{a(k)}=0\label{eq2}
\ee
where, $\tilde{a}(k) = (1/\sqrt{2\pi})\sum_{n} a_{n} \exp(-i k n)$. The equation for the stationary $mth$ mode $\tilde{a(k)}=e^{(i \beta^{(m)}z})\ u^{(m)}(k)$ is
\be
\tilde{u}^{(m)}= e^{(i/F) (\beta^{(m)}-2 \sin(k))}.
\ee
From the recurrence relation for the Bessel function, we obtain
$\beta^{(m)}=m\ F$ (Wannier-Stark ladder), and 
\begin{eqnarray}
u_{n}^{m} &=& {1\over{2 \pi}} \int_{-\pi}^{\pi} e^{(i k(n-m) + (2 i /F) \sin(k))}\nonumber\\&=&
             J_{n-m}(2/F).
\end{eqnarray}
where $J_{n}(x)$ is the Bessel function of the first kind. The reason for this behavior stems from an initial increase in linear momentum in the presence of the external field until the wavevector reaches the boundary of the first Brillouin zone, reentering the zone by the opposite edge. This oscillation in reciprocal space leads to the oscillation in real space.  

{\em Fractional effects.}\ Let us now add the effects of fractionality. Equation (\ref{eq1}) can be written as
\be 
i {d\over{d z}} a_{n}(z)+ n F a_{n}(z) + 2 a_{n}+(\Delta_{n})^{s} a_{n}=0 \label{eq2}
\ee
where $(\Delta_{n})^{s}$ is the discrete Laplacian $(\Delta_{n})^{s} a_{n} = a_{n+1}-2 a_{n}+a_{n-1}$. We proceed now to replace $(\Delta_{n})^{s}$ by its fractional form in Eq.(\ref{eq1}). The discrete fractional 1D laplacian is given in closed form by\cite{discrete Laplacian}
\be
(\Delta_{n})^{s} q_{n}=\sum_{m\neq n}K^{s}(n-m)(q_{m}-q_{n}), \hspace{1cm}0<s<1
\ee
where $s$ is the fractional exponent and
%%%%%%%%%%%%%%%%%%%%%%%%%%%%%%%%%%%%%%%%%%%%%%%%%%%%%%%%%%%%%
\begin{figure*}[t]
\includegraphics[scale=0.3]{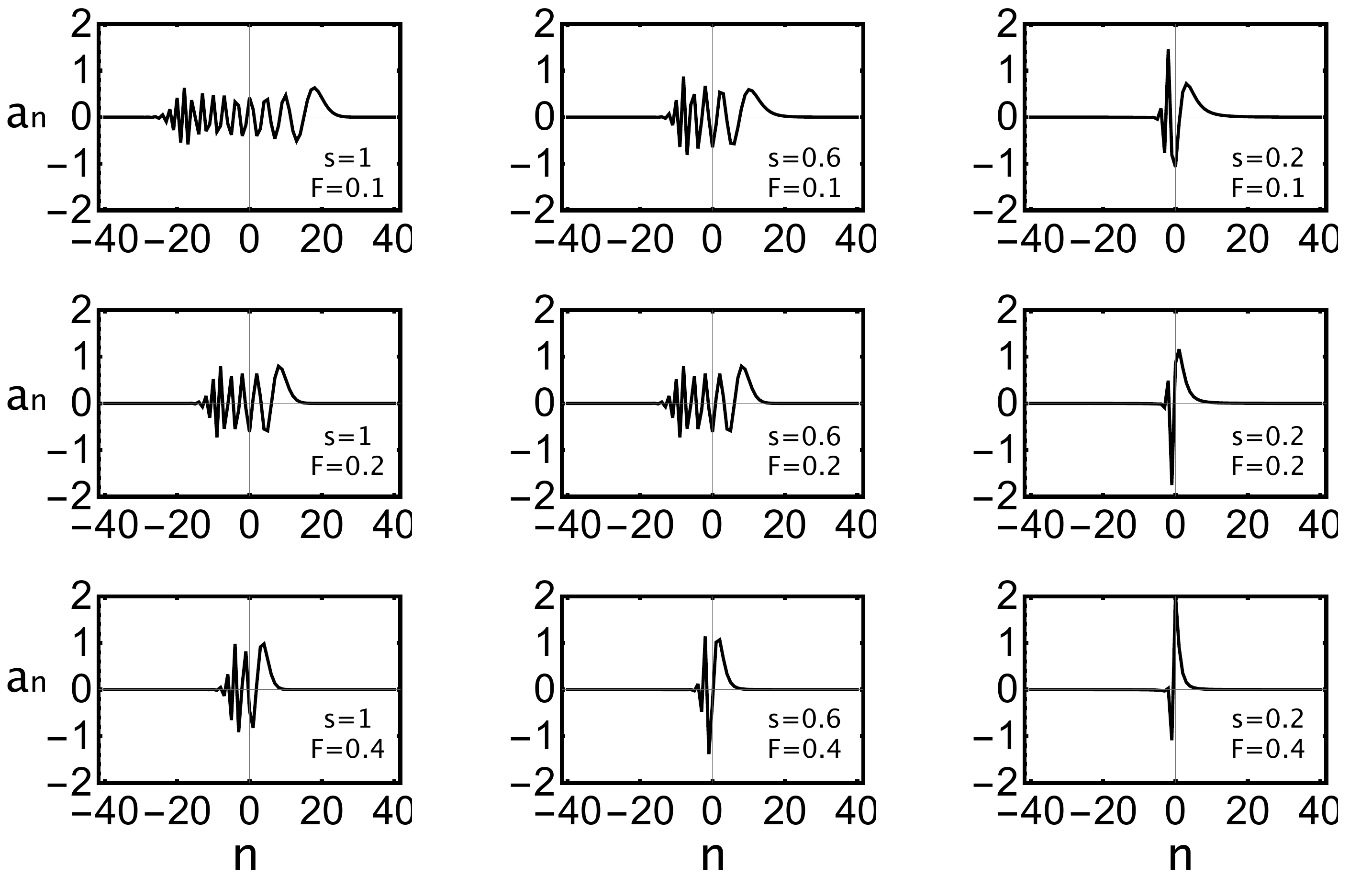}
\caption{Stationary mode $m=0$ for different fractional exponents and field strengths. }
\label{fig1}
\end{figure*}
%%%%%%%%%%%%%%%%%%%%%%%%%%%%%%%%%%%%%%%%%%%%%%%%%%%%%%%%%%%%%%
%%%%%%%%%%%%%%%%%%%%%%%%%%%%%%%%%%%%%%%%%%%%%%%%%%%%%%%%%%%%%
\begin{figure}
\includegraphics[scale=0.3]{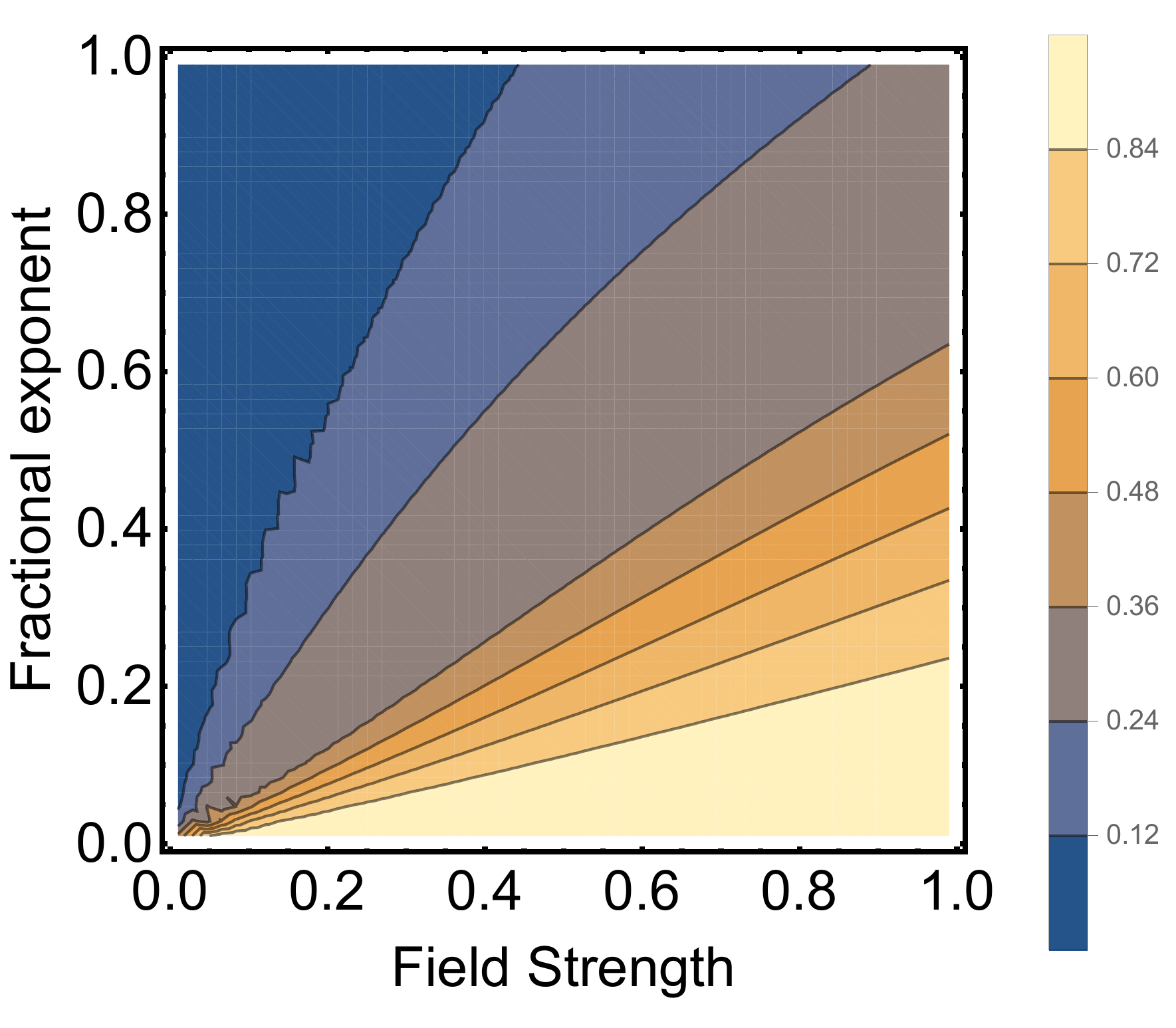}
\caption{Inverse participation ratio $R$ for the $m=0$ stationary mode for different fractional exponents $s$ and external field strengths $F$. }
\label{fig2}
\end{figure}
%%%%%%%%%%%%%%%%%%%%%%%%%%%%%%%%%%%%%%%%%%%%%%%%%%%%%%%%%%%%%%
\be
K^{s}(m) = {4^{s} \Gamma((1/2)+s)\over{\sqrt{\pi}|\Gamma(-s)|}}\  {\Gamma(|m|-s)\over{\Gamma(|m|+1+s)}}.\label{K}
\ee
and $\Gamma(x)$ is the Gamma function.
Equation (\ref{eq1}) becomes
\begin{eqnarray}
i {d\over{d z}} a_{n}(z)+ (2-\sum_{q\neq0} K^{s}(q))a_{n}+&& \nonumber\\
+n F a_{n}(z) + \sum_{q\neq0} K(q) a_{q+n}=0\label{eq3}
\end{eqnarray}
or, in wavector space
\be
i {d\over{d z}}\tilde{a} + i F {d\over{d k}}\tilde{a}+2 \sum_{q=1}^{\infty} K^{s}(q) \cos(q k) \tilde{a}=0\label{eq9}
\ee
The equation for the $mth$ stationary mode $\tilde{a}^{m}(z)=\tilde{u}^{m}(k)\ \exp{(i \beta_{m} z)}$ is 
\be
\tilde{u}^{m}(k)=e^{-(i/F)(\beta_{m} k -2\sum_{q=1}^{\infty}K^{s}(q) q^{-1}\sin(q k))}\ \ \ \ \ 
\ee
Because of periodicity, $\tilde{u}^{m}(-\pi)=\tilde{u}^{m}(\pi)$, this implies
$\beta_{m} = m F$, i.e., the Wannier-Stark ladder independent of the fractional exponent $s$. From this, we obtain in real space
\be
u^{m}_{n}={1\over{2 \pi}} \int_{-\pi}^{\pi} e^{(i k (n-m)-(2 i/F)\sum_{q=1}^{\infty} q^{-1} K^{s}(q) \sin(q k)  )} dk
\ee
or,
\be
u^{m}_{n}={1\over{\pi}} \int_{0}^{\pi} \cos\left( (k(n-m))-{2\over{F}}\sum_{q=1}^{\infty} {K^{s}(q)\over{q}}\sin(q k) \right)\label{eq5}.
\ee

We could further re-express the last term in Eq.(\ref{eq5})using

\begin{eqnarray}
\sum_{q=1}^{\infty} {K^{s}(q)\over{q}}\sin(q k)&=&{i\over{\sqrt{\pi}}}2^{2 s-1}e^{-i k}s \Gamma(s+{1\over{2}})\nonumber\\
& &\times \{\  H(\{1,1,1-s\},\{2,2+s\},e^{-i k})-\nonumber\\
& & e^{2 i k}\  H(\{1,1,1-s\},\{2,2+s\},e^{i k})\  \}\nonumber,
\end{eqnarray}
where $H(\{a\},\{b\},z)$ is the regularized hypergeometric function. But we choose to stick to Eq.(\ref{eq5}) since is more useful when exploring various limits.

Figure 1 shows some mode profiles for different fractional exponents $s$ and different field strengths $F$. For a fixed $s$, an increase in $F$ tends to localize the profile, while for a fixed field, a decrease in fractional exponent also tends to localize the mode. The spatial extent of a mode $u^{m}_{n}$ can be estimated by means of the Inverse Participation Ratio $R$, defined as 
\be
R^{m}(F,s)={\sum_{n} |u^{m}_{n}(F,s)|^4\over{(\sum_{n} |u^{m}_{n}(F,s)|^2}\ )^2}
\ee
thus, for a completely localized mode $u^m_{n}\sim \delta_{n,0}$, and $R^{m}\sim 1$, while a completely delocalized profile, $u^{m}_{n}\sim cte.$, leads to $R^{m}\sim 1/N$, where $N$ is the number of sites. Figure 2 shows $R^{0}(F,s)$ as a function of the external field strength and the fractional exponent, using a 2D (F,s) phase diagram. We see that for a fixed external field, a decrease in the fractional exponent increases $R$, signaling a tendency to localization. For a fixed fractional exponent, an increase in field strength also increases $R$, signaling localization as well.  

{\em Fractional dynamics.}\ Let us now compute the fractional dynamics of an initially completely localized excitation. We change variables in Eq.(\ref{eq9}) $ \tau = k - F z$. This leads to
\be
{d\over{d z}} \log(\tilde{a}) = 2 i \sum_{q=1}^{\infty} K^{s}(q) \cos(q (\tau + F z))
\ee
that is,
\be
\tilde{a} = e^{-(2 i/F)\sum_{q} {K^{s}\over{q}}(\sin(q (k-F z))-\sin(q k))}
\ee

After Fourier inverting, we obtain $a_{n}(z)$:
\be
a_{n}(z) = {1\over{(2 \pi)}}\int_{-\pi}^{\pi} e^{   
i k n -{2 i\over{F}}
\sum_{q}{K^s(q)\over{q}}
[\sin(q (k-F z))-\sin(q k)]]}\ dk
\ee
where, $a_{n}(0)=\delta_{n 0}$. The motion is periodic with period $T=2 \pi/F$, independent of the fractional exponent. Figure 3 shows some examples of the fractional dynamics for a fixed external field $F=0.2$ and several exponents $s$.
%%%%%%%%%%%%%%%%%%%%%%%%%%%%%%%%%
\begin{figure}[t]
 \includegraphics[scale=0.225]{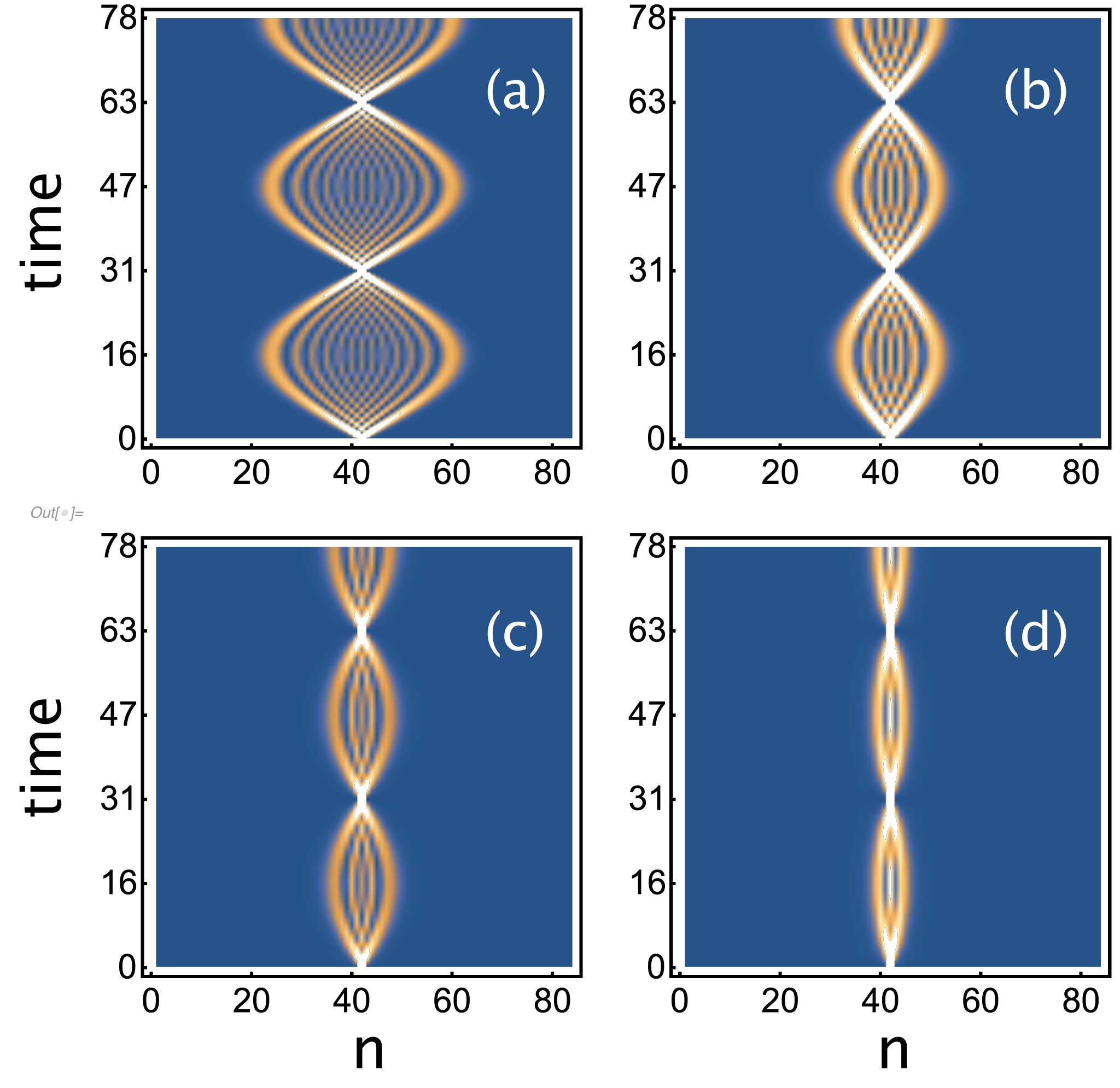}
  \caption{Propagation dynamics of a localized initial excitation, for $F=0.2$ and (a) $s=1.0$, (b) $s=0.6$, (c) $s=0.4$ (d) $s=0.2$}.
  \label{fig3}
\end{figure}
%%%%%%%%%%%%%%%%%%%%%%%%%%%%%%%%
\newline
As $s$ is decreased from $s=1$ (nonfractional case), the dynamic profiles narrow more and more, converging to a completely localized profile, as a consequence of $K^{s}(q)\rightarrow s/q$ as $s\rightarrow 0$. Results for the dynamical evolution of an initial Gaussian beam $a_{n}(0)=\exp(-(n-n_{c})^2/2)$ are shown in Fig. 4. As $s$ is decreased from the non-fractional limit ($s=1$), the width of the periodic profile also narrows, converging to the initial profile as $s\rightarrow 0$.
%%%%%%%%%%%%%%%%%%%%%%%%%%%%%%%%%
\begin{figure}[t]
 \includegraphics[scale=0.2]{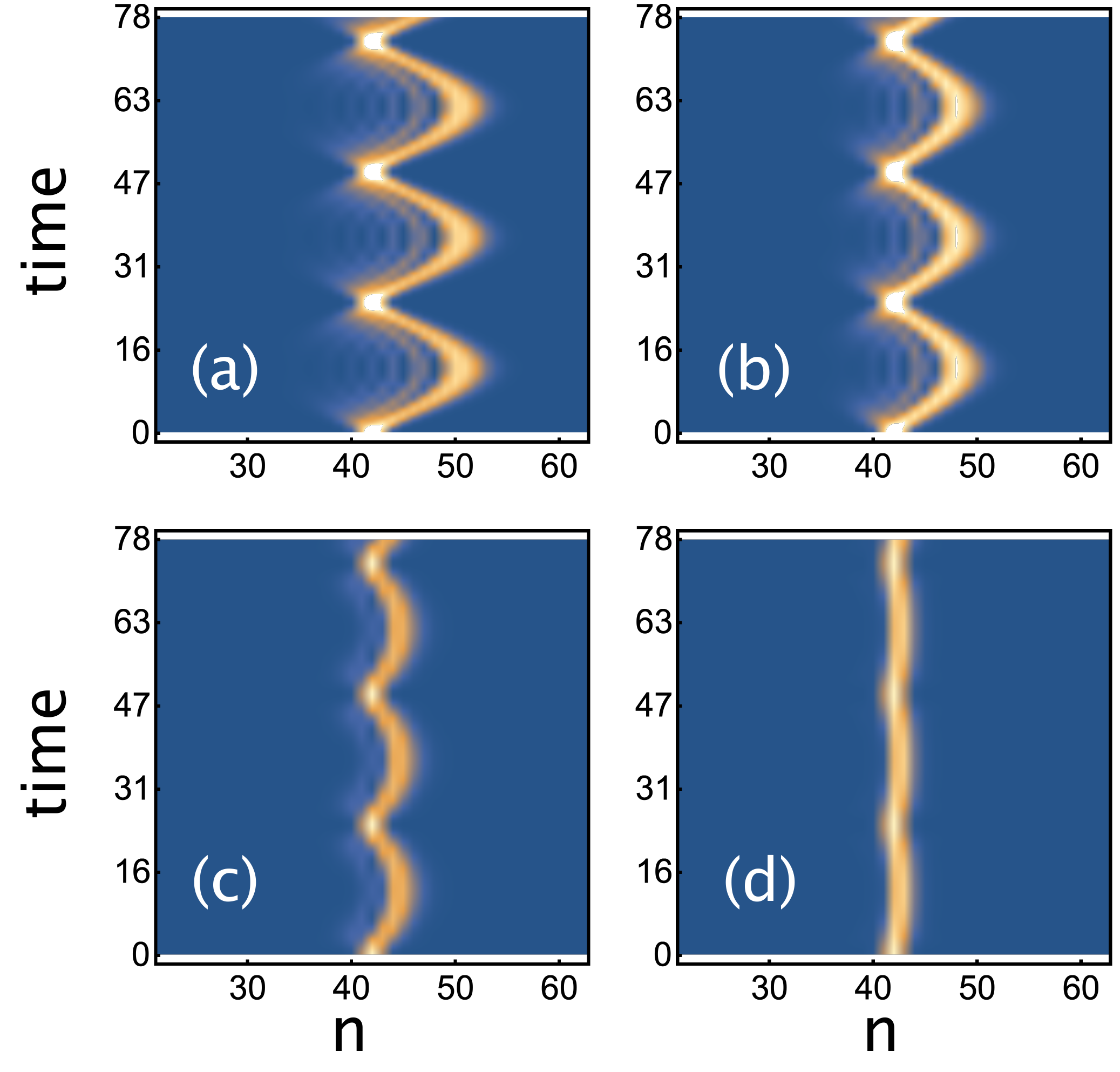}
  \caption{Propagation dynamics of a wide gaussian beam, for $F=0.4, n_{c}=42$ and (a) $s=1.0$, (b) $s=0.8$, (c) $s=0.4$ (d) $s=0.1$}.
  \label{fig4}
\end{figure}
%%%%%%%%%%%%%%%%%%%%%%%%%%%%%%%%

{\em Nonlinear effects}\ Finally, let us briefly look at the previous phenomenology when we incorporate some nonlinear effects. For coupled electron-phonon systems,  in the semiclassical limit, a common form of nonlinearity is the cubic (Kerr) nonlinearity, where the term $\chi |a_{n}(z)|^2 a_{n}(z)$
%%%%%%%%%%%%%%%%%%%%%%%%%%%%%%%%%
\begin{figure}[t]
 \includegraphics[scale=0.25]{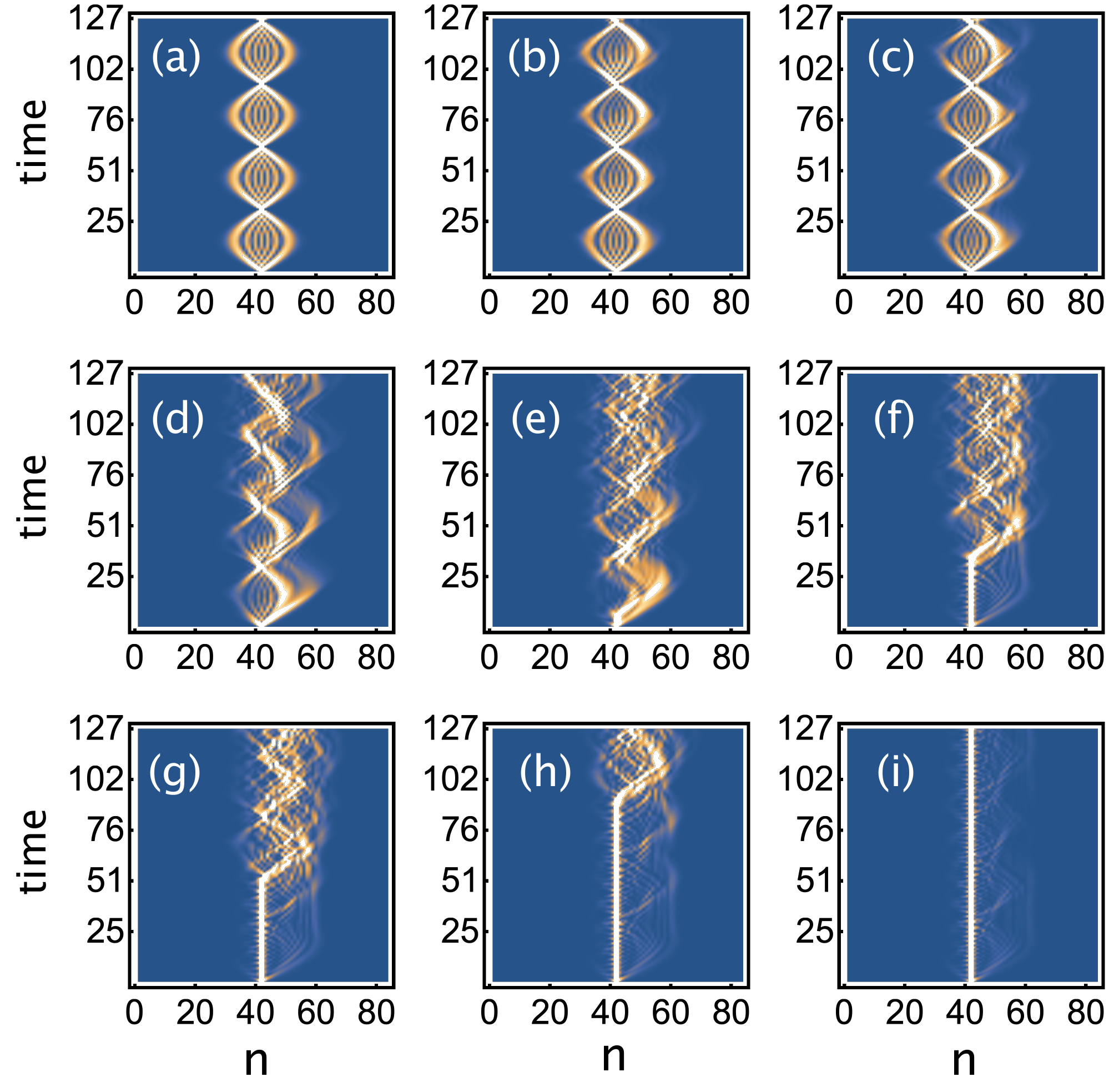}
  \caption{ Propagation dynamics of a localized initial excitation in the presence of nonlinearity. $s=0.6,F=0.2$ and (a) $\chi=0$, (b) $\chi=0.5$, (c) $\chi=1.0$, (d) $\chi=2.0$, (e) $\chi=3.0$, (f) $\chi=3.25$, (g) $\chi=3.3$, (h) $\chi=3.4$,(i) $\chi=3.5$ }.
  \label{fig5}
\end{figure}
%%%%%%%%%%%%%%%%%%%%%%%%%%%%%%%%
is added to Eq.(\ref{eq3}), which now reads
\begin{eqnarray}
i {d\over{d z}} a_{n}(z)+ (2-\sum_{q\neq 0} K^{s}(q))a_{n}(z)+
n F a_{n}(z) + &&\nonumber\\+\sum_{q\neq0} K(q) a_{q+n}(z)+\chi |a_{n}(z)|^2 a_{n}(z) =0.\label{ultima}
\end{eqnarray}

Figure \ref{fig5} shows some propagations of an initially excited site with fixed values of $s$ and $F$ and for different increasing values of the nonlinear parameter $\chi$. As can be appreciated, there is a competition between the tendency to Bloch oscillate and the trapping typical of the Kerr effect. At large values of $\chi$, the self-trapping dominates at finite times.

{\em Conclusions.}\ We have explored the influence of fractionality on the well-known Bloch oscillation (BO) phenomenology for an excitation propagating in a 1D periodic tight-binding lattice in the presence of an external constant field. Fractionality was introduced by replacing the standard discrete Laplacian by its fractional form, obtained recently in closed form for a 1D geometry\cite{discrete Laplacian}. The eigenmodes and eigenvalues were ultimately computed in closed form in terms of hypergeometric functions. A phase diagram for the inverse participation ratio in fractionality-external force space was computed in closed form, revealing a tendency towards localization with the external force and with the fractional exponent's decrease, away from the standard case. As the exponent decreases away from the non-fractional case, the BO profile decreases its width. In fact, in the limit of zero fractional exponent, the width of the BOs shrinks to zero in the case of an initially completely localized excitation. For a narrow Gaussian beam, the initial profile is also recovered at $s\rightarrow 0$.
The system also displayed a Wannier-Stark ladder with period $2 \pi/F$ , independent of the amount of fractionality. Since fractionality gives rise to a long-range coupling kernel among sites, this might suggest that the same independence is true for a more general model with an arbitrary coupling range. In fact, this behavior has already been noted for BO in the presence of first and second-nearest neighbors\cite{Adame}.   
When nonlinearity is introduced into the above picture as a cubic (Kerr) term, its trapping tendency competes with the BO, obliterating them in the limit of large nonlinearity.
Experimental realization of the fractional optics for fractional Schr\"{o}dinger-like equations has been proposed\cite{longhi} by means of a fiber-cavity setup, in which effective fractional group-velocity dispersion is implemented by means of a specially designed phase plate.

\vspace{1cm}
{\bf Acknowledgments}\\
\vspace{0.5cm}

This work was supported by Vicerrectoria de Investigacion y Desarrollo (VID) de la Universidad de Chile, grant ELN10/24.
\vspace{0.5cm}

{\bf Declaration of competing interest}\\

The authors declare that they have no known competing financial interests or personal relationships that could have appeared to influence the work reported in this paper.
\vspace{0.5cm}

{\bf Data availability}\\
No data was used for the research described in this article.

\end{document}